\documentclass[fleqn,usenatbib]{mnras}
\usepackage{newtxtext,newtxmath}
\usepackage[T1]{fontenc}
\usepackage{ae,aecompl}
\usepackage{graphicx}    
\usepackage{subfig}
\usepackage{ulem}
\usepackage{booktabs}

\title[Dark matter in NM and NS]{Effects of dark matter on the nuclear and neutron star matter}
\author[H. C. Das et al.]{
    H. C. Das$^{1,2}$
    \thanks{E-mail: harish.d@iopb.res.in},
    Ankit Kumar$^{1,2}$,
    Bharat Kumar$^{3}$,
    S. K. Biswal$^{4}$,  \newauthor
    Takashi Nakatsukasa$^{3}$,
    Ang Li$^{4}$,
    S. K. Patra$^{1,2}$,
    \\
    $^{1}$ Institute of Physics, Sachivalya Marg, Bhubaneswar-751005, India\\
    $^{2}$ Homi Bhabha National Institute, Training School Complex, 
    Anushakti Nagar, Mumbai 400085, India\\
    $^{3}$ Center for Computational Sciences, University of Tsukuba, Tsukuba 305-8577, Japan\\
    $^{4}$ Department of Astronomy, Xiamen University, Xiamen 361005, P. R. China}
    
\begin{document}
    \maketitle
    \date{\today}
    \begin{abstract}
        We study the dark matter effects on the nuclear matter parameters characterising the equation of states of super dense neutron-rich nucleonic-matter. The observables of the nuclear matter, i.e. incompressibility, symmetry energy and its higher-order derivatives in the presence dark matter for symmetric and asymmetric nuclear matter are analysed with the help of an extended relativistic mean-field model. The calculations are also extended to $\beta$-stable matter to explore the properties of the neutron star. We analyse the dark matter effects on symmetric nuclear matter, pure neutron matter and neutron star using NL3, G3 and IOPB-I forces. The binding energy per particle and pressure are calculated with and without considering the dark matter interaction with the nuclear matter systems. The influences of dark matter are also analysed on the symmetry energy and its different coefficients. The incompressibility and the skewness parameters are affected considerably due to the presence of dark matter in the nuclear matter medium. We extend the calculations to the neutron star and find its mass, radius and the moment of inertia for static and rotating neutron star with and without dark matter contribution. The mass of the rotating neutron star is considerably changing due to rapid rotation with the frequency in the mass-shedding limit. The effects of dark matter are found to be important for some of the nuclear matter parameters, which are crucial for the properties of astrophysical objects. 
    \end{abstract}
    \section{Introduction}\label{intro}
    It is well known that our Universe has only $\sim$ 6\% visible matter and the remaining $\sim$ 94\% is considered to be dark matter (DM) and dark energy. Zwicky estimated the total mass of the Universe \citep{Zwicky_2009, Kouvaris_2010} and found that something is missing, which are termed as DM ($\sim$ 26\%) and dark energy ($\sim$ 68\%). Many theoretical, experimental and observational efforts have been put to know the mystery of DM and dark energy. Several DM candidates are hypothesized, like weakly interacting massive particles (WIMPs) \citep{Kouvaris_2011,Quddus_2019}, feebly interacting massive  particles (FIMPs) \citep{Bernal_2017,Hall_2010}, Neutralino \citep{Hooper_2004,Han_2014,Das_2019} and axions \citep{Duffy_2009} etc. The WIMPs are expected to produce in the early hot Universe and annihilate in pairs, and these are the thermal relics of the Universe \citep{Ruppin_2014}. The WIMPs might have decayed in the dense region of the Universe to yield standard model (SM) particles, gamma rays, leptons and neutrinos. Many experiments have already been performed to find out the direct and indirect consequences of DM. The direct experimental searches like DAMA \citep{Bernabei_2008,Bernabei_2010}, Xenon \citep{Angle_2008} and CDMS \citep{CDMS_2010} are set up to find the cross-section between WIMPs and nucleons. The indirect detection experiments like Fermi large area telescopes and imaging air Cherenkov telescopes have also established \citep{Conrad_2014}. In addition to that, the effects of DM on compact stars such as neutron star (NS) and white dwarfs have been studied with different DM models \citep{Kouvaris_2008, Bertone_2008}. For example, the self-annihilating DM inside the NS heat the stars, and it would affect the cooling properties of compact stars \citep{Kouvaris_2008, Bhat_2019}. On the other hand, non-self-annihilating DM is accumulated inside the stars and affects the stellar structure \citep{De_Lavallaz_2010, Ciarcelluti_2011}. In this paper, we consider fermionic DM interacting with nucleonic matter via the Higgs portal mechanism and constraints the nuclear matter (NM) and NS properties through DM parameters. \\
    
    To understand equation of state (EoS) of NS matter, it is imperative to analyze the NM parameters at different proton-neutron compositions ($\alpha=\frac{\rho_n-\rho_p}{\rho_n+\rho_p})$, where $\rho_n$ and $\rho_p$ are the neutron and proton densities respectively. The NM parameters, such as binding energy per particle (BE/A), incompressibility ($K$), symmetry energy ($S$) and its derivatives ($L$-slope parameter, $K_{sym}$-isovector incompressibility and $Q_{sym}$-skewness parameter) are the key quantities for the study of an EoS. The NSs are the extreme object with a high degree of density and isospin asymmetry. Hence it is interesting to study the NM parameters at different conditions, from low to high density and at different asymmetric factor $\alpha=$ 0 to 1 in the presence of DM.
    In Ref. \citep{Alam_2016}, it was shown that the linear combinations of the isoscalar and isovector NM parameters are strongly correlated with NS radii over a wide range of NS mass. These correlations are particularly important for the canonical mass 1.4$M_\odot$ of the NS. With the help of GW170817 observation, a similar better correlation exists between the tidal deformability $\Lambda$ and the Love number $k_2$ with the linear combination of the $M_0$ and the curvature of the symmetry energy $K_{sym,0}$ at saturation density \citep{Tuhin_2018,Zack_2019}. Also, recently it is reported by various authors \citep{Sandin_2009,Kouvaris_2010,De_Lavallaz_2010,Ciarcelluti_2011,Leung_2011,AngLi_2012,Panotopoulos_2017,Ellis_2018,Bhat_2019,Das_2019,Ivanytskyi_2019,Quddus_2019} that the NS core contains an admixture of DM including many exotic baryonic species.\\
    
    The internal structure of a NS is not well known till now, so we believe many phenomena like quark deconfinement \citep{Collins_1975,Orsaria_2014,Mellinger_2017}, kaon condensation \citep{Kaplan_1998,NKGk1_1998,NKGk2_1999,Pal_2000,Gupta_2012}, phase transition \citep{NKGfp_1992,Sharma_2007} and hyperons production  \citep{Ambartsumyan_1960,NKGh_1985,Schaffner_1996,Schulze_2006,Apo_2010,Bipasa_2014,Biswal_2016,Fortin_2017,Bhuyan_2017,Biswalaip_2019,Biswal_2019} are occur inside the star. The EoS plays a vital role to predict all the star parameters such as mass ($M$), radius ($R$), tidal deformability ($\Lambda$) and moment of inertial ($I$) of the NS. Many theoretical and observational studies have been devoted to constraint these parameters. The binary NS merger event GW170817 \citep{Abbott_2017,Abbott_2018} provides a strong constraint on the EoS. The recently reported massive NS (PSR J0740+6620) \citep{Cromartie_2019} with the mass of $2.14^{+0.20}_{-0.18}$  $M_\odot$ within the $95.4\%$ confidence limits, also puts a strong constraint on the nature of EoS.\\
    
    Recently the simultaneous measurements of the $M$ and $R$ for NS are done by the NASA Neutron star Interior Composition ExploreR (NICER) \citep{Miller_2019,Riley_2019,Bogdanov_2019,Bilous_2019,Raaijmakers_2019,Guillot_2019}, which constraint the EoS. For that, we extend our calculations for rotating NS to measure the $M$, $R$ and $I$ in the presence of DM. The theoretical observations allow the Keplerian frequency of the rotating NS is more than 2000 $Hz$, but two fastest pulsar detected have frequencies 716 $Hz$ \citep{Hessels_2006} and 1122 $Hz$ \citep{Kaaret_2007}. Many calculations related to the Keplerian frequencies \citep{Stergioulas_2003,Dhiman_2007,Jha_2008,Krastev_2008,Haensel_2009,Sharma_2009,Koliogiannis_2020} are devoted to fix the frequency range within this limit. In this work, we want to study the effect of mass-shedding frequency on the mass of the DM admixture NS, which can be used to constraint the EoS.\\
    
    The EoS, which is the main input to the Tolman-Oppenheimer-Volkoff (TOV) equations \citep{TOV1,TOV2} determine the stable configurations of a static NS, are constructed in several ways. The non-relativistic formalism with various Skyrme parametrizations \citep{Skyrme_1956,Skyrme_1958,Vautherian_1972,Chabanta_1998,Brown_1998,Stone_2007,Dutra_2012, Gogny_1980} and three-body potential of Akmal-Pandheripande \citep{Akmal_1998} are very successful in describing the nuclear EoS, including the NS. The relativistic mean-field (RMF) model, which gives a good description not only explains well the finite nuclei in the $\beta$-stability line but also reproduce the experimental data for exotic and superheavy nuclei \citep{Rashdan_2001,Bhuyan_2012,Bhuyan_2018,Kumar_2018}. In the present paper, we use the extended RMF (E-RMF) models for the study of effects of DM on NM and NS properties with the well known NL3 \citep{Lalazissis_1997}, G3 \citep{Kumar_2017} and IOPB-I \citep{Kumar_2018} parameter sets.\\
    
    The effects of the DM on the NS have already discussed in some recent works, for example, Panotopoulos et al. \citep{Panotopoulos_2017}, Das et al. \citep{Das_2019} and Quddus et al. \citep{Quddus_2019}. But our analysis gives a better and wider platform to discuss the DM effects on the bulk properties of the NM and the NS. Here, we briefly describe their works. Panotopouls et al. \citep{Panotopoulos_2017} have calculated the NS EoS with simple $\sigma$-$\omega$ model \citep{Walecka_74}, and they have added the DM with static NS to find its $M$ and $R$. They assumed that the DM particles interact with nucleons via SM Higgs. They considered Neutralino as a DM candidate, which has mass 200 GeV and Higgs mass is 125 GeV.  Das et al. \citep{Das_2019} have taken the same RMF model but with NL3 parameter set \citep{Lalazissis_1997} and calculate the EoS, $M$, $R$ and $\Lambda$ of the static NS. Quddus et al. \citep{Quddus_2019} have taken the E-RMF formalism, which is suitable for constraints on the properties of the NS in the presence of WIMP dark matter. The main difference between earlier two work and Quddus et al. is that they have taken light Higgs mass 40 GeV and DM mass up to 8 GeV. In the present calculations, we take three non-linear parameter sets, whose nuclear matter properties cover a wide range. Though simple $\sigma$-$\omega$ model gives a qualitative picture of the RMF model, still many important information are missing in the linear $\sigma$-$\omega$ model. But E-RMF calculation covers a full set of non-linear model contain various important interaction terms as in Refs. \citep{Kumar_2017,Kumar_2018}. We add the DM with the NM and calculate NM parameters like BE/A, $K$, $S$, $L$, $K_{sym}$ and $Q_{sym}$ for the whole density range. Finally, we apply the $\beta$-equilibrium condition to the NM EoS. When we add the DM to the NM, we follow the same formalism by the Panotopouls et al. \cite{Panotopoulos_2017} and Das et al. \citep{Quddus_2019}. Then we calculate the $M$, $R$ and $I$ both for static and rotating NS and we compare both the results. \\
    
    The paper is organised as follows: the formalism used in this work is presented in Sec. \ref{TF}. In Sub-Sec. \ref{RMF}, we explain the basic formalism of RMF model using NL3 and the recently developed G3 and IOPB-I forces for the calculations of nucleonic EoS. In Sub-Sec. \ref{NDM}, we take the interaction of DM with NM and calculate the EoS of nucleons with DM. In Sub-Sec. \ref{SNMP}, we calculate different parameters of NM. The Sub-Sec. \ref{NSP} the calculation of EoS of NS using $\beta$-equilibrium and charge-neutrality conditions. In Sub-Sec. \ref{RNS}, the observables of the NS are calculated like $M$, $R$, $I$ etc. for static NS (SNS) and rotating NS (RNS). The results and discussions are detailed in Sec. \ref{R&D}. Finally, summary and our concluding remarks are outlined in Sec. \ref{CONCLU}. 
    \section{Theoretical Framework}
    \label{TF}
    From the last four decades, the RMF approaches are extremely popular to describe finite and infinite NM properties. The Lagrangian is constructed by taking the interaction of few numbers of mesons with nucleons and their self and cross-couplings. The parameters are constructed, taking into account the experimental data of few finite nuclei, saturation properties of the infinite NM and NS properties.
    \subsection{Construction of RMF approach to find EoS for nucleons}\label{RMF}
    The RMF Lagrangian is built from the interaction of mesons-nucleons and their
    self ($\Phi^3$,$\Phi^4$, $W^4$) and cross-couplings ($\Phi^2-W^2$, $R^2-W^2$, $\Phi-W^2$ and $\Phi-R^2$) of the mesons fields $\Phi$, $W$, $R$, $D$. Where $\Phi$, $W$, $R$ and $D$ are the redefined fields for $\sigma$, $\omega$, $\rho$ and $\delta$ mesons as
    $\Phi = g_s\sigma^0$, $W = g_\omega \omega^0$, $R$ = g$_\rho\vec{\rho}$ $^0$ and $D=g_\delta\delta^0$ respectively. The RMF Lagrangian is discussed in these Refs.  \citep{Miller_1972,Serot_1986,Furn_1987,Reinhard_1988,Frun_1997,Kumar_2017,Kumar_2018}. The energy density (${\cal{E}}_{nucl.}$) and pressure ( $P_{nucl.}$) for a nucleon-meson interacting system are given as \citep{Kumar_2018,Quddus_2019}
    \begin{eqnarray}
    {\cal{E}}_{nucl.}&=& \frac{\gamma}{(2\pi)^{3}}\sum_{i=p,n}\int_0^{k_i} d^{3}k E_{i}^\star (k_i)+\rho_bW +\frac{1}{2}\rho_{3}R\nonumber\\
    &&
    +\frac{ m_{s}^2\Phi^{2}}{g_{s}^2}\Bigg(\frac{1}{2}+\frac{\kappa_{3}}{3!}
    \frac{\Phi }{M_{nucl.}} + \frac{\kappa_4}{4!}\frac{\Phi^2}{M_{nucl.}^2}\Bigg)-\frac{1}{4!}\frac{\zeta_{0}W^{4}}
    {g_{\omega}^2}
    \nonumber\\
    &&
    -\frac{1}{2}m_{\omega}^2\frac{W^{2}}{g_{\omega}^2}\Bigg(1+\eta_{1}\frac{\Phi}{M_{nucl.}}+\frac{\eta_{2}}{2}\frac{\Phi ^2}{M_{nucl.}^2}\Bigg)
    \nonumber\\
    &&
    -\Lambda_{\omega}  (R^{2}\times W^{2})-\frac{1}{2}\Bigg(1+\frac{\eta_{\rho}\Phi}{M_{nucl.}}\Bigg)\frac{m_{\rho}^2}{g_{\rho}^2}R^{2}
    \nonumber\\ 
    &&
    +\frac{1}{2}\frac{m_{\delta}^2}{g_{\delta}^{2}}D^{2},
    \label{eq1}
    \end{eqnarray}
    \noindent
    \begin{eqnarray}
    P_{nucl.} &=&  \frac{\gamma}{3 (2\pi)^{3}}\sum_{i=p,n}\int_0^{k_i} d^{3}k \frac{k^2}{E_{i}^\star (k_i)}+\frac{1}{4!}\frac{\zeta_{0}W^{4}}{g_{\omega}^2}\nonumber\\
    &&
    -\frac{ m_{s}^2\Phi^{2}}{g_{s}^2}\Bigg(\frac{1}{2}+\frac{\kappa_{3}}{3!}
    \frac{\Phi }{M_{nucl.}}+ \frac{\kappa_4}{4!}\frac{\Phi^2}{M_{nucl.}^2}\Bigg)\nonumber\\
    &&
    +\frac{1}{2}m_{\omega}^2\frac{W^{2}}{g_{\omega}^2}\Bigg(1+\eta_{1}\frac{\Phi}{M_{nucl.}}+\frac{\eta_{2}}{2}\frac{\Phi ^2}{M_{nucl.}^2}\Bigg)
    \nonumber\\
    &&
    +\Lambda_{\omega} (R^{2}\times W^{2})+\frac{1}{2}\Bigg(1+\frac{\eta_{\rho}\Phi}{M_{nucl.}}\Bigg)\frac{m_{\rho}^2}{g_{\rho}^2}R^{2}
    \nonumber\\
    &&
    -\frac{1}{2}\frac{m_{\delta}^2}{g_{\delta}^{2}}D^{2} \nonumber.\\
    \label{eq2}
    \end{eqnarray}
    The energy density of the nucleon in the meson medium is $E_{i}^\star(k_i)$=$\sqrt {k_i^2+{M_{i}^\star}^2}$, where $M_i^\star$ is the effective mass calculated in Eq. (\ref{eq6}) and $k_i$ is the momentum of the nucleon, where i = p, n. The $\rho_b$ and $\rho_3$ in Eq. (\ref{eq1}) are the baryonic and iso-vector density respectively. $\gamma$ is the spin degeneracy factor which is equal to 2 for individual nucleons. $M_{nucl.}$ is the mass of the nucleon which is 939 MeV.
    \subsection{Interaction between nucleons and DM candidates in NM}\label{NDM}
    It is a well known fact that the NS is rotating along with the galaxy. The DM halo in the Universe is also rotating so that the
    DM particles accreted mostly in the NS core due to its very high gravitational field and high baryon density  \citep{Goldman_1989,Kouvaris_2008,Xiang_2014,Das_2019}. When DM particles interact with nucleons, it loses energy and helps in the cooling of the NS \citep{Bhat_2019}. The cooling of the NS are detailed in these Refs. \citep{Gendin_2001,Page_2004,Yakovlev_2004,Yakovlev_2005,Yakovlev_2010}. The amount of DM inside the NS depends on the evolution time of the NS in the Universe. In this context, we consider the Neutralino \citep{Martin_1998,Panotopoulos_2017} as a fermionic DM candidate which interact with nucleon via SM Higgs.The interaction Lagrangian of DM and nucleons is given as \citep{Panotopoulos_2017,Quddus_2019,Das_2019}
    \begin{eqnarray}
    {\cal{L}} & = & {\cal{L}}_{nucl.} + \bar \chi \left[ i \gamma^\mu \partial_\mu - M_\chi + y h \right] \chi +  \frac{1}{2}\partial_\mu h \partial^\mu h  \nonumber\\
    & &
    - \frac{1}{2} M_h^2 h^2 + f \frac{M_{nucl.}}{v} \bar \varphi h \varphi , 
    \label{eq3}
    \end{eqnarray}
    where ${\cal{L}}_{nucl.}$ is the nucleon-mesons Lagrangian and $\varphi$ and $\chi$ are the nucleonic and DM wave functions respectively. We take the mass of Neutralino ($M_\chi$) is 200 GeV, and the coupling constants between DM and SM Higgs is $y$, which can be found in the large Higgs mixing angle limit. Since the Neutralino is the super symmetric particle, it has the various gauge coupling constants in the electroweak sector of the standard model \citep{Martin_1998}. So depending on the different parameters, the values of $y$ is  in the range 0.001 -- 0.1. Thus we take the value of $y$ = 0.07 in our calculations. The Higgs field directly couples to the nucleons with Yukawa interaction $f \frac{M_{nucl.}}{v}$, where $f$ is proton-Higgs form factor. The detailed analytical expression for $f$ can be found in \citep{Cline_2013}. In lattice calculations \citep{Alarc_n_2012,Young_2013}, we can consider the value of $f=0.35$, which is agreement with \citep{Cline_2013}. The Higgs mass is $M_h=$ 125 GeV. The vacuum expectation value ($v$) of Higgs is 246 GeV. From the Lagrangian in Eq. (\ref{eq3}), we get the total energy density (${\cal{E}}$) and pressure ($P$) for NM with DM given as \citep{Panotopoulos_2017,Das_2019,Quddus_2019} 
    \begin{eqnarray}
    {\cal{E}}& = &  {\cal{E}}_{nucl.} + \frac{2}{(2\pi)^{3}}\int_0^{k_f^{DM}} d^{3}k \sqrt{k^2 + (M_\chi^\star)^2 } 
    \nonumber\\
    & &
    + \frac{1}{2}M_h^2 h_0^2 ,
    \label{etot}
    \label{eq4}
    \end{eqnarray}
    \begin{eqnarray}
    P& = &  P_{nucl.} + \frac{2}{3(2\pi)^{3}}\int_0^{k_f^{DM}} \frac{d^{3}k \hspace{1mm}k^2} {\sqrt{k^2 + (M_\chi^\star)^2}} 
    \nonumber\\
    & &
    - \frac{1}{2}M_h^2 h_0^2 ,
    \label{ptot}
    \label{eq5}
    \end{eqnarray} 
    where $k_f^{DM}$ is the DM Fermi momentum. We consider the baryon density inside NS is 1000 times larger than the DM density, this imply that $M_{\chi}/M$=1/6 \citep{Panotopoulos_2017,Das_2019}, where $M$ is the mass of the NS. One can get $k_f^{DM}$ is $\sim$ 0.03 GeV \citep{Das_2019}. So that we vary $k_f^{DM}$ from 0 -- 0.06 GeV. The effective masses of nucleon and DM are given as
    \begin{eqnarray}
    M_i^\star &=& M_{nucl.}+ g_\sigma \sigma_0 \mp g_\delta \delta_0 - \frac{f M_{nucl.}}{v}h_0, 
    \nonumber\\
    M_\chi^\star &=& M_\chi -y h_0,
    \label{eq6}
    \end{eqnarray}
    where the $\sigma_0$, $\delta_0$ and $h_0$ are the meson field equations of $\sigma$, $\delta$ and Higgs respectively and  these are obtained by  applying mean field approximations, which are given in Ref. \citep{Das_2019}. 
    The DM density $\rho_{\chi}$ is
    \begin{equation}
    \rho_\chi =  \frac{\gamma}{(2 \pi)^3}\int_0^{k_f^{\rm DM}}\frac{M_\chi^\star}{\sqrt{M_\chi^\star{^2}+ k^2}} d^3k \label{eq7},
    \end{equation}
    \subsection{NM Parameters}\label{SNMP}
    The calculations of NM properties need the energy density and pressure as a function of baryonic density. The energy density ${\cal{E}}$ can be expanded in a Taylor series in terms of $\alpha$ \citep{Horowitz_2014,Baldo_2016,Kumar_2018}.
    \begin{equation}
    {\cal{E}}(\rho,\alpha) = {\cal{E}}(\rho,\alpha=0)+S(\rho) \alpha^2+ {\cal{O}}(\alpha^4),
    \label{eq8}
    \end{equation}
    where ${\cal{E}}(\rho,\alpha = 0)$ is the energy of symmetric NM , $\rho$ is the baryonic density and $S(\rho)$ is the symmetry energy, which is defined as 
    \begin{equation}
    S(\rho) = \frac{1}{2}\Bigg(\frac{\partial^2{\cal{E}}}{\partial\alpha^2}\Bigg)_{\alpha=0}
    \label{eq9}
    \end{equation}
    The symmetry energy is  also written as the energy difference between pure neutron matter (PNM) and symmetric NM (SNM) or vice-versa through parabolic approximation, i.e.
    \begin{equation}
    S(\rho) = \frac{{\cal{E}}(\rho,\alpha = 1)}{\rho}-\frac{{\cal{E}}(\rho,\alpha = 0)}{\rho}.
    \label{eq10}
    \end{equation}
    Although, the value of symmetry energy is fairly known at the saturation density ($\rho_0$), its density dependence nature is not well known. The behavior of $S(\rho)$ in high density, both qualitatively and quantitatively shows a great diversion depending on the model used \citep{BaoLi_2019}. Similar to the BE/A, the $S(\rho)$ can also be expressed in a leptodermous expansion near the NM saturation density. The analytical expression  of density dependence symmetry energy is written as \citep{Matsui_1981,Kubis_1997,MCentelles_2001,Chen_2014,Kumar_2018}:
    \begin{equation}
    S(\rho) = J+L\zeta+\frac{1}{2}K_{sym}\zeta^2+\frac{1}{6}Q_{sym}\zeta^3+{\cal{O}}(\zeta^4),
    \label{eq11}
    \end{equation}
    where $\zeta$=$\frac{\rho-\rho_0}{3\rho_0}$, $J$ = $S(\rho_0$) and the parameters like slope ($L$), curvature ($K_{sym}$) and skewness ($Q_{sym}$) of $S(\rho$) are
    \begin{eqnarray}
    L=3\rho\frac{\partial S(\rho)}{\partial\rho}, \label{eq12}\\
    K_{sym}=9\rho^2\frac{\partial^2 S(\rho)}{\partial\rho^2}, \label{13}\\
    Q_{sym}=27\rho^3\frac{\partial^3 S(\rho)}{\partial\rho^3}. \label{eq14}
    \end{eqnarray}
    The NM incompressibility ($K$) is defined as \citep{Chen_2014}
    \begin{eqnarray}
    K=9\rho^2\frac{\partial}{\partial \rho}\Bigg(\frac{P}{\rho^2}\Bigg).
    \label{eq15}
    \end{eqnarray}
    To estimate both symmetry energy and its slope parameters at saturation density $\rho_0$, we use Eq.  (\ref{eq10}). Since the parameters $J$ and $L$ play important roles like, formation of clusters in finite nuclei and normal star, dynamic of heavy-ion collisions and cooling process of newly born NS. These parameters are also crucial for the study of phase transition (finite/infinite nuclear systems). Different approaches are available for the calculation of $J$ and $L$ including their correlations \citep{MCentelles_2009,Xu_2010,Fattoyev_2012,Steiner_2012,Newton_2012,Dutra_2012,Singh_2013}.
    \subsection{The EoS of NS }\label{NSP}
    In this section, we describe the NS EoS within a medium of nucleons, electrons and muons. In the NS, the neutron decays to proton, electron and anti-neutrino \citep{NKGb_1997,Quddus_2019,Bhat_2019}. There is also inverse $\beta$--decay to
    maintain the beta equilibrium and charge neutrality condition. This can be expressed as \citep{NKGb_1997}
    \begin{eqnarray}
    n \rightarrow p+e^-+\bar\nu, \nonumber\\ 
    p+e^-\rightarrow n+\nu .
    \label{eq16}
    \end{eqnarray}
    The stability of NSs is followed by $\beta$-equilibrium and charge-neutrality conditions as follow as   
    \begin{eqnarray}
    \mu_n &=& \mu_p +\mu_e,   \nonumber \\
    \mu_e &=& \mu_\mu,
    \label{eq17}
    \end{eqnarray}
    where, $\mu_n$, $\mu_p$, $\mu_e$, and $\mu_\mu$ are the chemical potentials of neutrons, protons, electrons, and muons, respectively, and the charge neutrality conditions is
    \begin{eqnarray}
    \rho_p = \rho_e +\rho_\mu. 
    \label{eq18}
    \end{eqnarray}
    The chemical potentials $\mu_n$, $\mu_p$, $\mu_e$, and $\mu_\mu$ are given by 
    \begin{equation}
    \mu_n = g_\omega \omega_0 + g_\rho \rho_0+\sqrt{k_n^2+ (M_n^\star)^2},
    \label{eq19}
    \end{equation} 
    \begin{equation}
    \mu_p =  g_\omega \omega_0 - g_\rho \rho_0+\sqrt{k_p^2+ (M_p^\star)^2},
    \label{eq20}
    \end{equation}
    \begin{equation}
    \mu_e = \sqrt{k_e^2+ m_e^2},
    \label{eq21}   
    \end{equation}
    \begin{equation}
    \mu_\mu = \sqrt{k_\mu^2+ m_\mu^2},
    \label{eq22}
    \end{equation}
    where $M_n^\star$ and $M_p^\star$ is the effective masses of neutron and proton respectively calculated in Eq. (\ref{eq6}). To find the particle fraction, we solve Eq. (\ref{eq17}) and (\ref{eq18}) followed by Eqs. (\ref{eq19} -- \ref{eq22}) in a  self-consistent way for a given baryon density. The total energy density and pressure of NS are given by,
    \begin{eqnarray}
    {\cal {E}}_{NS} &=& {\cal {E}} + {\cal {E}}_{l}, \nonumber \\ and \hspace{1cm}
    P_{NS} &=& P + P_l,
    \label{eq23}
    \end{eqnarray}
    where, 
    \begin{equation}
    {\cal {E}}_{l} = \sum_{l=e,\mu}\frac{2}{(2\pi)^{3}}\int_0^{k_l} d^{3}k \sqrt{k^2 + m_l^2 },
    \label{eq24}
    \end{equation}
    and
    \begin{equation}
    P_{l} = \sum_{l=e,\mu}\frac{2}{3(2\pi)^{3}}\int_0^{k_l} \frac{d^{3}k \hspace{0.2cm}k^2} {\sqrt{k^2 + m_l^2}}.
    \label{eq25}
    \end{equation}
    Where ${\cal{E}}_{l}$, $P_{l}$ and $k_l$ are the energy density, pressure and Fermi momentum for leptons respectively. The Eq. (\ref{eq23}) gives the total energy, pressure and number density of the NS. 
    \subsection{Observables of the NS}
    \label{RNS}
    In the Sec. \ref{intro}, we have already mention that NS is very complex structure and  therefore the detailed study require both theories of GR and dense matter. i.e NS forms a link between two fundamental theory in the modern physics and this connection already given by Einstein's field equation \citep{Einstein_1916}
    \begin{eqnarray}
    G^{\mu\nu}=R^{\mu\nu}-\frac{1}{2}g^{\mu\nu}R=8\pi T^{\mu\nu},
    \label{eq26}
    \end{eqnarray}
    where $G^{\mu\nu}$, $R^{\mu\nu}$, $g^{\mu\nu}$ and $R$ are the Einstein tensor, Ricci tensor, metric tensor and Ricci scalar respectively. The $T^{\mu\nu}$ is the energy-momentum tensor for perfect fluid is given as \citep{NKGb_1997,Krastev_2008}
    \begin{eqnarray}
    T^{\mu\nu}=({\cal E}_{NS}+P_{NS})u^{\mu}u^{\nu}+Pg^{\mu\nu},
    \label{eq27}
    \end{eqnarray}
    where ${\cal E}_{NS}$ and $P_{NS}$ are the energy density and pressure of the NS.  $u^{\mu}$ is the 4-velocity satisfying, $u^{\mu}u_{\mu}=-1$. The $T^{\mu\nu}$ directly depends on the EoS of the stellar matter in the form of $P_{NS}({\cal E}_{NS})$. To solve the Einstein's Eq. \ref{eq26}, First, we have to solve the short-range nuclear forces of many-body nuclear physics in a local inertial frame in which space-time is flat. Second, the long range force of the gravitational field which describe the curvature of space-time created by the massive objects \citep{Krastev_2008}. We calculate the many-body theory using RMF formalism in Sub-Sec. \ref{RMF}. But in this present Sub-Sec. we calculate the different observables like $M$, $R$ and $I$ etc. for the NS. In our whole calculations we use $G=c=1$. \\
    
    In case of static, spherically symmetric stars the metric is in the form of 
    \begin{eqnarray}
    ds^2= -e^{2\nu(r)}dt^2+e^{2\lambda(r)}dr^2+r^2d\theta^2+r^2sin^2\theta d\phi^2,
    \label{eq28}
    \end{eqnarray}
    where $r$, $\theta$ and $\phi$ are the spherical co-ordinates. $\nu(r)$, $\lambda(r)$ are the metric potential are given as  \citep{Krastev_2008}
    \begin{eqnarray}
    e^{2\lambda(r)} = [1-\gamma(r)]^{-1},
    \label{eq29}
    \end{eqnarray}
    \begin{eqnarray}
    e^{2\nu(r)}&=&e^{-2\lambda(r)} = [1-\gamma(r)], \qquad r>R_{star}
    \label{eq30}
    \end{eqnarray}
    with
    \begin{equation}\label{eq.31}
    \gamma(r)=\left\{
    \begin{array}{l l}
    \frac{2m(r)}{r}, & \quad \mbox{if $r<R_{star}$}\\\\
    \frac{2M}{r}, & \quad \mbox{if $r>R_{star}$}
    \end{array}
    \right.
    \end{equation}
    For static star, the Einstein's Eq. \ref{eq26} reducess to 
    \begin{eqnarray}
    \frac{dP_{NS}(r)}{dr}= - \frac{[P_{NS}(r)+{\cal{E}}_{NS}(r)][m(r)+4\pi r^3 P_{NS}(r)]}{r[r-2m(r)]},\label{eq32}
    \end{eqnarray}
    and 
    \begin{eqnarray}
    \frac{dm(r)}{dr}=4\pi r^2 {\cal{E}}_{NS}(r),\label{eq33}
    \end{eqnarray}
    where ${\cal{E}}_{NS}(r)$ and $P_{NS}(r)$ are the total energy density and pressure appearing in Eq. (\ref{eq23}) under the $\beta$-equilibrium condition. $m(r)$ is the gravitational mass, and $r$ is the radial parameter. These two coupled equations are solved to get the $M$ and $R$ of the SNS at certain central density. \\
    
    The metric of RNS in equilibrium is described by a stationary and axisymmetric metric of the form \citep{Stergioulas_2003} 
    \begin{eqnarray}
    ds^2= -e^{2\nu}dt^2+e^{2\psi}(d\phi - \omega dt^2)+e^{2\alpha}(r^2d\theta^2+ d\phi^2),
    \end{eqnarray}
    where $\nu$, $\psi$, $\omega$ and $\alpha$ are metric functions which depend on $r$ and $\theta$ only. The energy-momentum tensor same as in Eq. \ref{eq27}, but the 4-velocity is changed as follow as
    \begin{equation}
    u^{\mu}=\frac{e^{\mu}}{\sqrt{1-v^2}}(t^{\mu}+\Omega\phi^{\mu}),
    \end{equation} 
    and the spatial velocity, $v=(\Omega-\omega)e^{\psi-\nu}$ , where $\Omega$ is the angular velocity of the star. $t^{\mu}$ and $\phi^{\mu}$ are two killing vectors in the space-time associated with time and transnational symmetries.\\
    
    Many people have calculated the moment of inertia of the NS \citep{Stergioulas_2003,Jha_2008,Sharma_2009,Friedmanstergioulas_2013,Paschalidis_2017,Quddus_2019,Koliogiannis_2020}. The simple expression is $I=J/\Omega$, where $J$ is angular momentum and $\Omega$ is the angular velocity of the NS. The expression of $I$ of uniformly rotating NS with angular frequency $\omega$ is given as \citep{Lattimer_2000,Worley_2008}
    \begin{equation}
    I \approx \frac{8\pi}{3}\int_{0}^{R}({\cal E}_{NS}+P_{NS})\ e^{-\phi(r)}\Big[1-\frac{2m(r)}{r}\Big]^{-1}\frac{\Bar{\omega}}{\Omega}r^4 dr,
    \label{eq36}
    \end{equation}
    where the $\Bar{\omega}$ is the dragging angular velocity for a uniformly rotating star. The $\Bar{\omega}$ satisfying the boundary conditions are 
    \begin{equation}
    \Bar{\omega}(r=R)=1-\frac{2I}{R^3},\qquad \frac{d\Bar{\omega}}{dr}\big|_{r=0}=0 .
    \label{eq37}
    \end{equation}
    The Keplerian frequency of the NS also calculated in these Refs.  \citep{Komatsu1_1989,Komatsu2_1989,Stergioulas_2003,Dhiman_2007,Jha_2008,Krastev_2008,Worley_2008,Haensel_2009,Sharma_2009,Koliogiannis_2020} and the expression is given as 
    \begin{equation}
    \nu_k(M) \approx \chi \Big(\frac{M}{M_{\odot}}\Big)^{1/2}\Big(\frac{R}{10\ \ km}\Big)^{-3/2},
    \end{equation}
    where $M$ and $R$ are  the gravitational mass  and  radius of the RNS respectively. The value of $\chi$ is calculated from the fitting or empirically \citep{Lattimer_2004,Haensel_2009}. In Ref. \citep{Haensel_2009}, they have found the value of $\chi=1.08$ kHz empirically. For rotating NS, we calculate the $M$, $R$, $I$ and Keplerian frequency ($\nu_K$) using public-domain program, RNS code written by the Stergioulas \citep{NikolaosStergioulas_1999}.
    \section{Results and Discussions}\label{R&D}
    In this section, we present the calculated results for BE/A, $K$, $S$ and its derivatives for NM varying the  $k_f^{DM}$ at different $\alpha$. We extend the calculations to NSs and find the $M$, $R$ and $I$.
    \begin{table}
        \caption{The 3-parameter sets NL3 \citep{Lalazissis_1997} , G3 \citep{Kumar_2017} and  IOPB-I \citep{Kumar_2018} are listed. All the coupling constants are dimensionless.}
        \centering
        \scalebox{1.0}{
\begin{tabular}{cccccccccc}
\hline
\hline
\multicolumn{1}{c}{Parameter}
&\multicolumn{1}{c}{NL3}
&\multicolumn{1}{c}{G3}
&\multicolumn{1}{c}{IOPB-I}\\
\hline
$m_{\sigma}/M_{nucl.}$  &  0.541  &  0.559&0.533  \\
$m_{\omega}/M_{nucl.}$  &  0.833 &  0.832&0.833  \\
$m_{\rho}/M_{nucl.}$  &  0.812 &  0.820&0.812  \\
$m_{\delta}/M_{nucl.}$   & 0.0  &   1.043&0.0  \\
$g_{\sigma}/4 \pi$  &  0.813   &  0.782 &0.827 \\
$g_{\omega}/4 \pi$  &  1.024  &  0.923&1.062 \\
$g_{\rho}/4 \pi$  &  0.712 &  0.962 &0.885  \\
$g_{\delta}/4 \pi$  &  0.0  &  0.160& 0.0 \\
$k_{3} $   &  1.465  &    2.606 &1.496 \\
$k_{4}$  &  -5.688  & 1.694 &-2.932  \\
$\zeta_{0}$  &  0.0  &  1.010  &3.103  \\
$\eta_{1}$  &  0.0  &  0.424 &0.0  \\
$\eta_{2}$  &  0.0  &  0.114 &0.0  \\
$\eta_{\rho}$  &  0.0 &  0.645& 0.0  \\
$\Lambda_{\omega}$  &  0.0 &  0.038&0.024\\
\hline
\hline
    \end{tabular}}
    \label{PMT}
    \end{table}
    \subsection{NM Properties}\label{NM}
\begin{figure}
        \includegraphics[width=0.9\columnwidth]{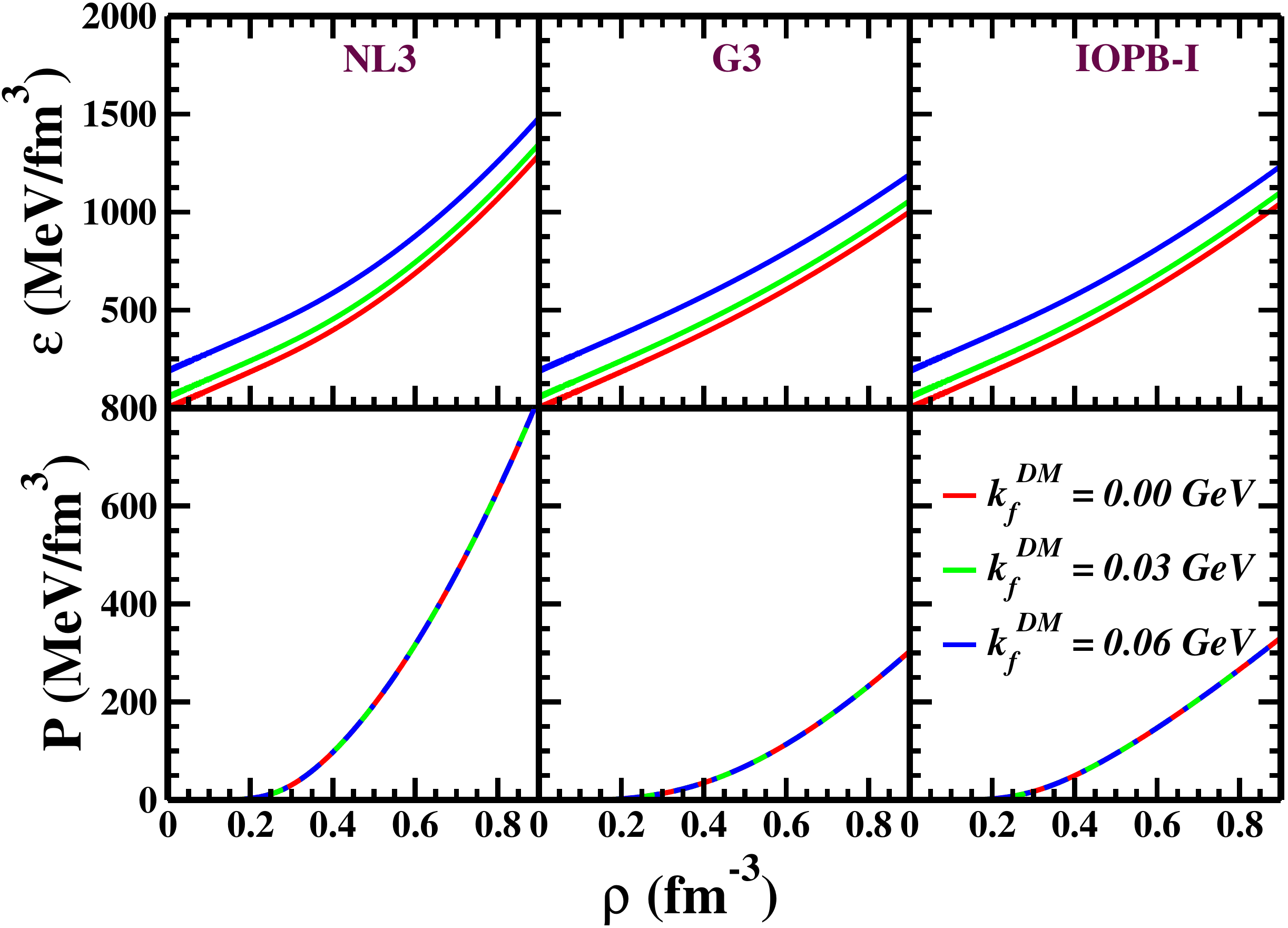}
        \caption{(colour online) The energy density ( Eq. \ref{eq4}) and pressure ( Eq. \ref{eq5}) for SNM with baryon density at $k_f^{DM}$= 0.0, 0.03 and 0.06 GeV.}
        \label{NMEoS}
    \end{figure}

\begin{figure}
        \centering
        \includegraphics[width=1.0\columnwidth]{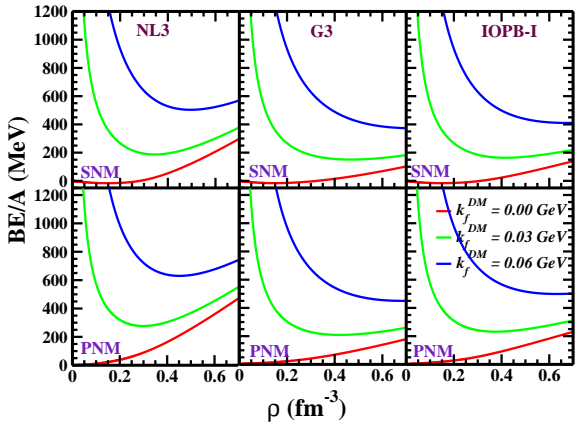}
        \caption{(color online) The BE/A of NM in the presence of DM at $k_f^{DM}$= 0.0, 0.03 and 0.06 GeV.}
        \label{BE}
    \end{figure}

    The $\cal E$ and $P$ are obtained using RMF approaches with DM \citep{Panotopoulos_2017,Quddus_2019,Das_2019} in Eqs. \ref{eq4} and \ref{eq5}. The ${\cal{E}}$ and $P$ as a function of baryon density $\rho$ are shown in Fig. \ref{NMEoS}. The RMF forces NL3 \citep{Lalazissis_1997}, G3 \citep{Kumar_2017} and IOPB-I \citep{Kumar_2018} are given in Table \ref{PMT}. Since, the $k_f^{DM}$ is not yet settled, we change its values for $k_f^{DM}=$ 0.0, 0.03 and 0.06 GeV and note down the variations as a function of $\rho$ in Fig. \ref{NMEoS}. We notice that the value of ${\cal{E}}$ changes significantly without affecting the $P$ (see Fig. \ref{NMEoS}).
    NL3 gives the stiffest EoS as compare to IOPB-I and G3 for the SNM case. Here also, G3 predicts the softest EoS, which is shown in the Fig. \ref{NMEoS}. Thus, the qualitative nature of the EoS is similar with and without the presence of DM as far as stiffness or softness is concerned. The BE/A is defined as $\frac{\cal{E}}{\rho}$-$M_{nucl.}$, where the $\cal{E}$ is the total energy density in Eq. (\ref{eq4}) and $\rho$ is the baryon density. The BE/A as a function of $\rho$ at different $k_f^{DM}$ is shown in Fig. \ref{BE} for both SNM and PNM. Here the effect of DM on BE/A is significant with respect to $k_f^{DM}$ for both SNM and PNM.\\

\begin{figure}
        \centering
        \includegraphics[width=1.0\columnwidth]{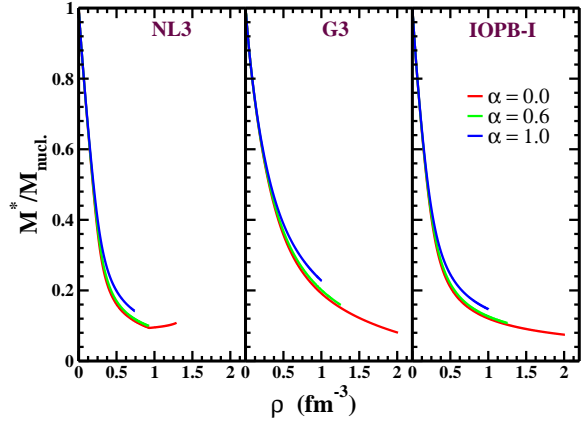}
        \caption{(colour online) The variation of effective mass in Eq. (\ref{eq6}) for different $\alpha$ with baryon density with $k_f^{DM}$ = 0.06 GeV.}
        \label{EFFM}
    \end{figure}

    The NS is mostly made of neutrons with a small fraction of protons, electrons and muons in the charge-neutral and $\beta$-equilibrium system. Thus to get an idea about the NM parameters, we check the variation of the effective mass ($M^\star$) with different $\alpha$. Here it is imperative to mention that $k_f^{DM}$ = 0 GeV means $\rho_{\chi}$ is zero, but the effect on $M^\star$ is very less due to non-zero Higgs-nucleon Yukawa coupling in Eq. (\ref{eq6}). The contribution of Higgs filed is very small ${\cal{O}}$($10^{-6}-10^{-8}$) even after varying  $k_f^{DM}$ to its maximum value. So that we plot the effective mass to mass ratio of the nucleon ($M^\star/M_{nucl.}$) as a function of $\rho$ for different $\alpha$, which is shown in Fig. \ref{EFFM}. In the presence of DM, the $M^\star/M_{nucl.}$ decreases with baryon density $\rho$, similar to the normal nuclear medium. As far as the neutron to proton ratio (N/Z) increases, the $M^\star/M_{nucl.}$ value goes on increasing mostly in the high-density region. However, there is practically no effect of $\alpha$ in the low-density region of the NM system.\\

\begin{figure}
        \centering
        \includegraphics[width=0.9\columnwidth]{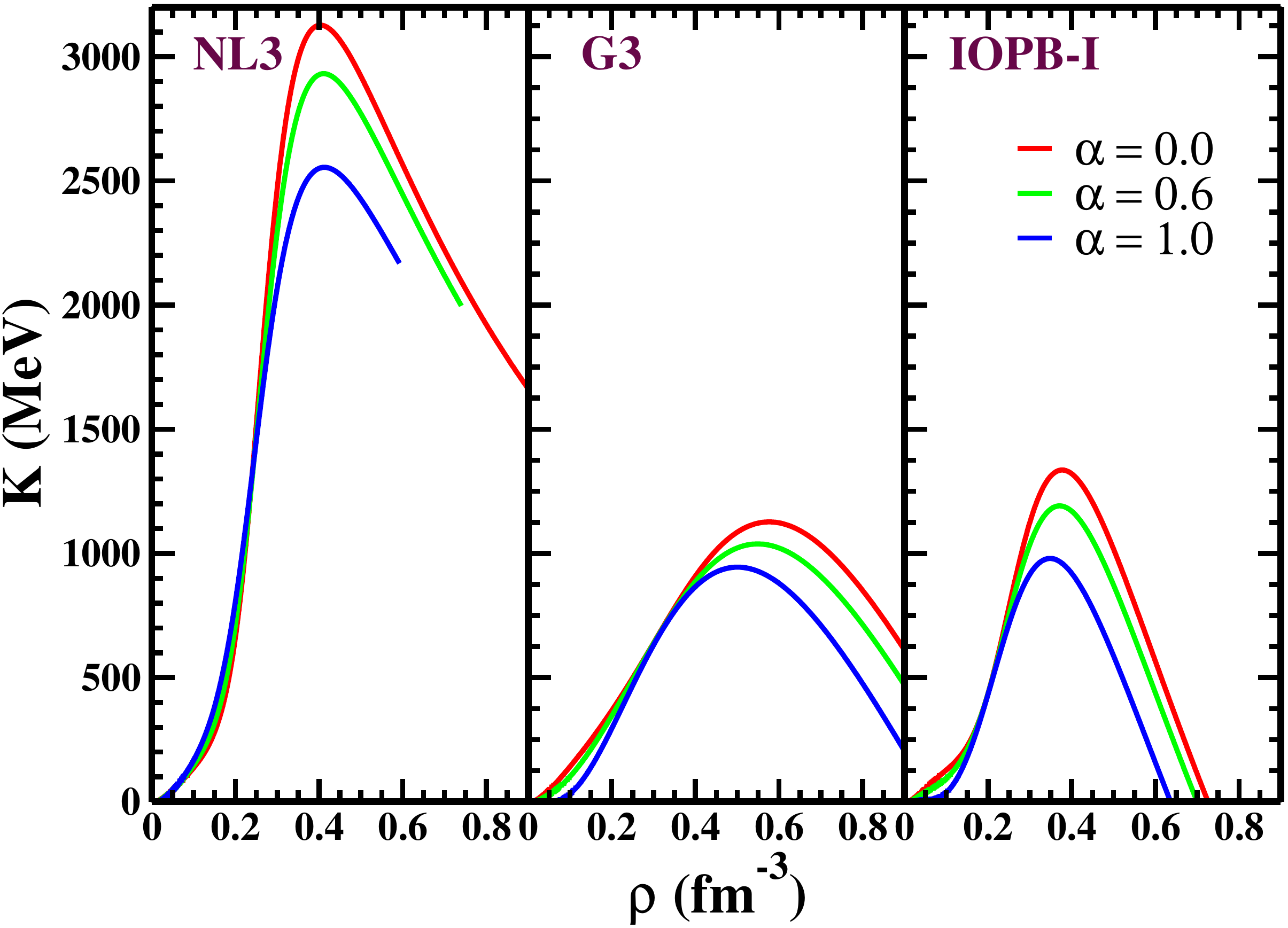}
        \caption{(colour online) The variation of incompressibility $K$ with different $\alpha$ as a function of baryon density $\rho$ with $k_f^{DM}$ = 0.06 GeV.}
        \label{ICOMP} 
    \end{figure}

    Another important NM parameter is $K$. This value tells us how much one can compress the NM system. It is a standard quantity at the saturation point. However, an astronomical object like the NS, its density varies from the centre of the star to the crust with a variation of $\rho$ from $\rho=10\rho_0$ to 0.1$\rho_0$ \citep{Lattimer_2004}. Thus, to achieve better knowledge on the compression mode or monopole vibration mode, we have to calculate the $K$ for all the density range of NM with different $\alpha$ including 0 and 1. Since we see the earlier case, DM does not affect on the pressure of either SNM or PNM also in NS (in sec. \ref{NS}), so DM doesn't affect the $K$ of NM. The variation of $K$ with baryon density for different $\alpha$ displayed in Fig. \ref{ICOMP}. One can see in Table \ref{NMT}, for SNM system the incompressibility at saturation $K_{\infty}$ are 271.38, 243.96 and 222.65 MeV for NL3, G3 and IOPB-I respectively. It is worthy of mentioning that the DM does not affect on the incompressibility. That means, the $K$ values remain unaffected with the variation of $k_f^{DM}$. On the other hand, substantial variation is seen with the different $\alpha$. We found that the value of $K$ increases initially up to a maximum and then gradually decreases, as shown in Fig. \ref{ICOMP}. The calculations also show that with increasing $\alpha$, the incompressibility decreases irrespective of the parameter sets. The values of $K$ for G3 and IOPB-I parameter sets lie in the region (except for NL3) given by the experimental value in Table \ref{NMT}. Since the NL3 gives very stiff EOS so that it's all NM parameters like $K$, $J$, $L$ etc. provide very large values and do not lie the region given in Table \ref{NMT}.\\

    The recent gravitational wave observation from the merger of two NSs, the GW170817 \citep{Abbott_2017,Abbott_2018}, constraints the upper limit on the tidal deformability $\Lambda$ and predicts a small radius. Also, the recent discovery of the three highly massive stars $\sim$ 2 $M_\odot$ \citep{Antoniadis_2013,Fonseca_2016,Cromartie_2019} predicts that the pressure in the inner core of the star is large, where the typical baryon number density quite high $\rho$ $>$ 3$\rho_0$ in this region. The pressure in the outer core of the massive NS is considered to be small in the density range 1 to 3 $\rho_0$ \citep{McLerran_2019}. Combining these observations of large masses and the smaller radii of the  massive NSs, one can infer that the causality \citep{Rhoades_1974,Bedaque_2015,Kojo_2015,Moustakidis_2017,McLerran_2019} of the NM inside the inner core of the NS can violate  \citep{McLerran_2019}. It is conjectured that the speed of the sound $C_s\leq c/\sqrt{3}$, where $C_s^2=\frac{\partial P}{\partial {\cal{E}}}$ with $C_s$ and $c$ are the speed of the sound and light, respectively.
    \begin{figure}
        \centering
        \includegraphics[width=0.9\columnwidth]{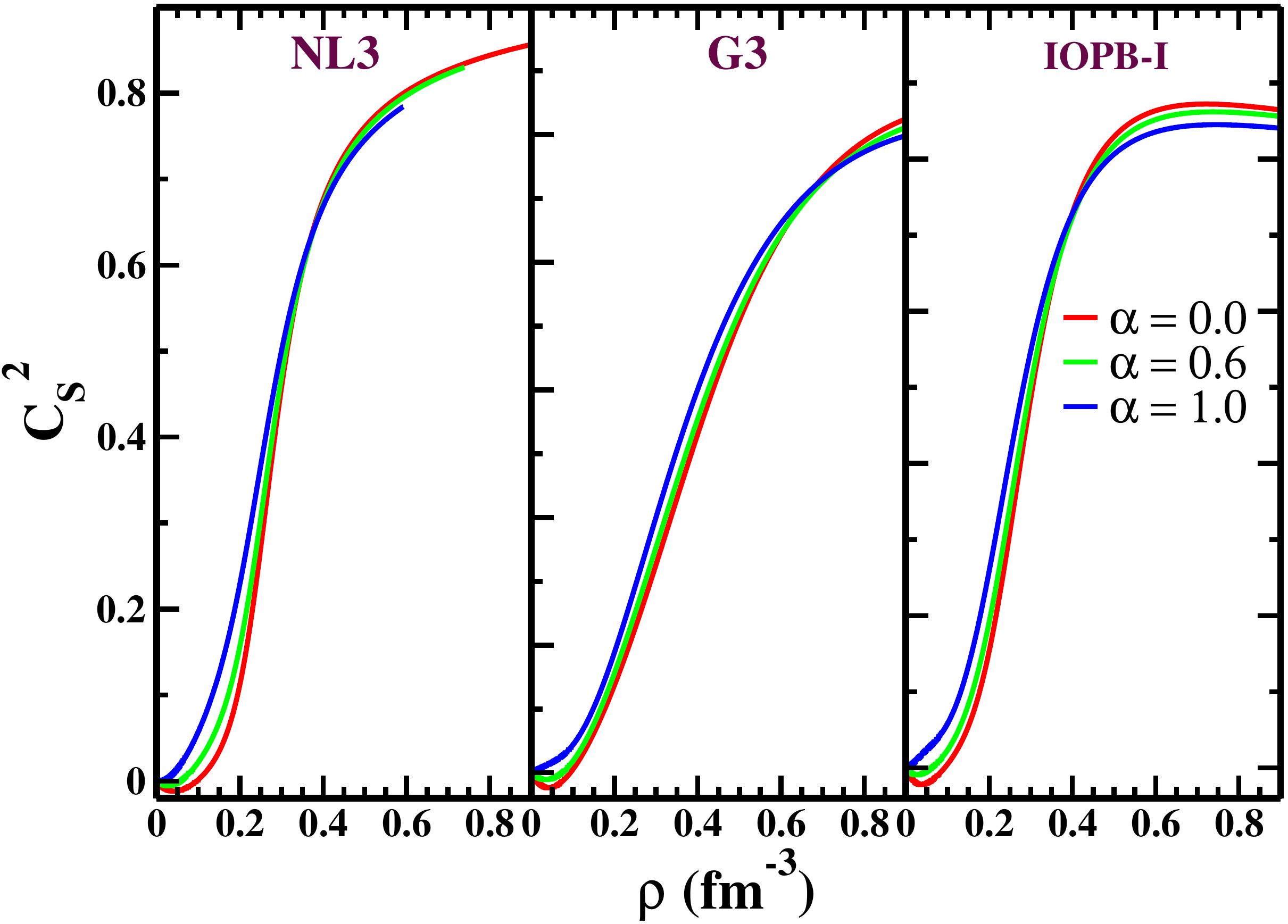}
        \caption{(colour online) The variation of speed of the sound with baryon density for NL3, G3 and IOPB-I parametrizations at different $\alpha$ with $k_f^{DM}$ = 0.06 GeV.}
        \label{VEL}
    \end{figure}
    To see the causality condition for the NM case with an admixture of DM, we plot $C_s^2$ as a function of baryon number density $\rho$ in Fig. \ref{VEL} at different $\alpha$ for NL3, G3 and IOPB-I parameter sets. We find that the $C_s^2$ increases approximately up to 0.8 fm$^{-3}$, then it is constant for high density regions. It is clear from the values of $C_s^2$ that the causality remains intact for a wide range of density for all three parameter sets, as shown in Fig. \ref{VEL}. \\
    
    The symmetry energy $S$ and its coefficients $L$, $K_{sym}$ and $Q_{sym}$ are defined in Eqs. (\ref{eq12} -- \ref{eq14}), play a crucial role in the EoS for symmetric and asymmetric NM.
    \begin{figure}
        \centering
        \includegraphics[width=0.9\columnwidth]{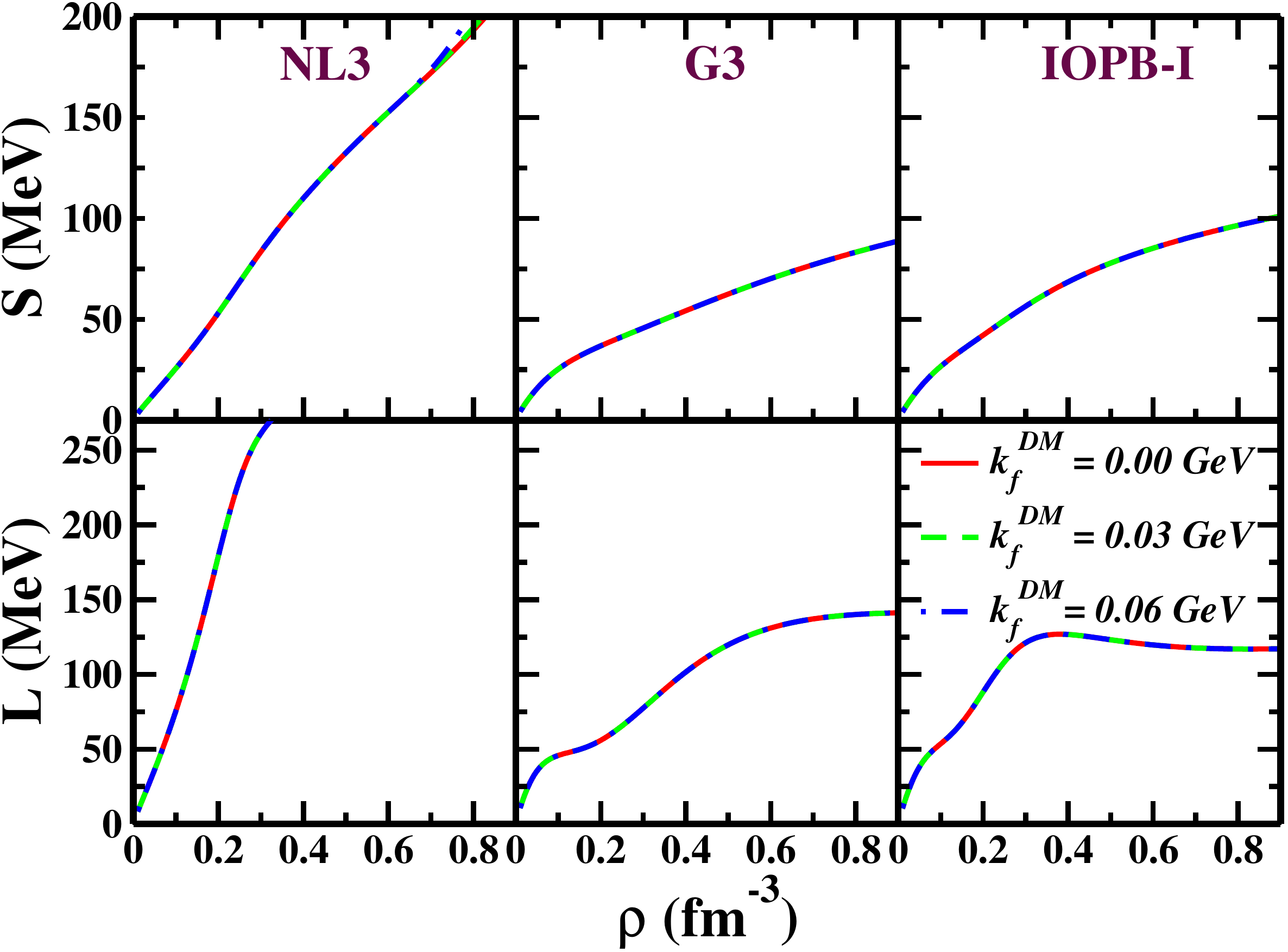}
        \caption{(colour online) The symmetry energy $S$ and its slope parameter $L$ are plotted with the varying $k_f^{DM}$ as a function of with $\rho$. We found almost the same results ( Table \ref{NMT} ) with and without DM for all the three-parameter sets.}
        \label{SYMM}
    \end{figure}
    As we have mentioned earlier, these parameters are important quantities to determine the nature of the EoS. Just after the supernova explosion, its remnants, which lead towards the formation of a NS, is in a high temperature ($\sim 200$ MeV) state \citep{Gendin_2001,Page_2004,Yakovlev_2004,Yakovlev_2005,Yakovlev_2010}. Soon after the formation of the NS, it starts cooling via the direct urca processes to attain the stable charge neutrality and $\beta$-equilibrium condition. The dynamical process of NS cooling is affected heavily by these NM parameters. Thus, it is very much intuitive to study these parameters in more rigorously.
    \begin{table*}
        \centering
        \captionsetup{width=1.0\textwidth}
        \caption{The nuclear matter properties such as BE/A, symmetry energy and its derivatives etc. are given at saturation density for three-parameters sets with k$_f^{DM}$ = 0 GeV (without DM), 0.03 GeV and 0.06 GeV respectively for SNM. Similarly the effective mass ($M^\star/M_{nucl}$), incompressibility ($K$) varying with $\alpha$ = 0, 0.6, 1.0 at $\rho_0$ (not in the variations of k$_f^{DM}$) in last three-rows. The empirical/expt. value for $\rho_0$, BE/A, $J$, $L$, $K_{sym}$ and $K$ are also given at saturation density.}
        \renewcommand{\tabcolsep}{0.15cm}
        \renewcommand{\arraystretch}{1.5}
        \begin{tabular}{lllllllllll}
            \hline\hline
            Parameters & \multicolumn{3}{l}{\hspace{1.2cm} NL3} & \multicolumn{3}{l}{ \hspace{1.4cm}G3} & \multicolumn{3}{l}{\hspace{1.2cm}IOPB-I}&
            \multicolumn{1}{l}{\hspace{1.0cm}Empirical/Expt. value}\\ 
            \cmidrule(lr){2-4}\cmidrule(lr){5-7}\cmidrule(lr){8-10}  
            &  0.0   &  0.03     &  0.06     &   0.0    &  0.03     & 0.06      &   0.0    &  0.03  & 0.06          \\ \hline
            $\rho_0$ (fm$^{-3}$)        &  0.148  &  0.148    &  0.148    &   0.148   &  0.148    & 0.148     &   0.149   &  0.149    &  0.149 & \hspace{1.2cm} 0.148 -- 0.185 \citep{Bethe_1971}   \\ 
            BE/A (MeV)                    & -16.35  &  143.95   &  1266.11  &   -16.02  &  143.28   & 1266.44   &   -16.10  &  143.09   & 1257.51  & \hspace{1.0cm} -15 -- -17 MeV \citep{Bethe_1971} \\ 
            $J$ (MeV)                     &  37.43  &  38.36    &  38.36    &   31.84   &  31.62    & 31.62     &   33.30   &  34.45    &  34.45   & \hspace{1.0cm} 30.20 -- 33.70 MeV \citep{Danielewicz_2014} \\ 
           $L$ (MeV)                     &  118.65 & 121.44    &  121.45   &   49.31   &  49.64    & 49.64     &   63.58   &  67.16    &  67.67   & \hspace{1.0cm} 35.00 -- 70.00 MeV \citep{Danielewicz_2014}   \\ 
            $K_{sym}$ (MeV)             &  101.34 & 101.05    &  100.32   &   -106.07 &  -110.38  & -111.10 & -37.09  &  -45.94   & -46.67   & \hspace{1.0cm} -174 -- -31 MeV \quad \citep{Zimmerman_2020}  \\ 
            $Q_{sym}$ (MeV)             &  177.90 & 115.56    &  531.30   &   915.47  &  929.67   & 1345.40   &   868.45  &  927.84   &  1343.58  & \hspace{1.2 cm}----------------  \\ \hline \hline 
            $\alpha$ =                  &  \hspace{0.2cm}0       & \hspace{0.1cm}0.6      & \hspace{0.1cm} 1.0      &   \hspace{0.2cm} 0       & \hspace{0.1cm} 0.6      &  \hspace{0.1cm}1.0      &  \hspace{0.2cm} 0       &  \hspace{0.1cm}0.6      &  \hspace{0.1cm}1.0   &   \\ \hline 
            $M^\star/M_{nucl.}$                        &  0.595   & 0.596    &  0.606    &  0.699    &  0.700    &  0.704    &   0.593   &  0.599    & 0.604  & \hspace{1.0cm}---------------  \\ \hline
            $K$ (MeV)                     &  271.38  & 312.45   &  372.13   &  243.96   &  206.88   & 133.04    &   222.65  &  204.00   & 176.28  & \hspace{1.0cm}220 -- 260 MeV \quad \citep{Stone_2014}    \\ \hline\hline
        \end{tabular}
        \label{NMT}
    \end{table*}
    The $S$ and its $L$-coefficient for the whole density range for all three parameter sets NL3, G3 and IOPB-I with different $k_f^{DM}$ are displayed in Fig. \ref{SYMM}. The effect of DM on $J$ and $L$ is very small, and it is difficult to notice in the figure. To have the knowledge, we summarised the results at the saturation point in Table \ref{NMT}. For example, with NL3 set, $S(\rho_0$) = $J$ = 37.43 MeV for PNM and it increases to a value $J$ = 38.36 MeV with DM. Similarly, $L$ = 118.65 MeV for without DM, $L$ = 121.44 and 121.45 MeV in the presence of DM for various DM momentum $k_f^{DM}$. For other sets, one can see the variation in Table \ref{NMT}. This is due to the fact that the effect of DM does not change NM asymmetry to a significant extent. A careful inspection of Fig. \ref{SYMM} and Table \ref{NMT}, it is clear that $S$ and $L$ are force-dependent. It is maximum for NL3 and minimum for G3 sets. The experimental value of $J$ and $L$ is given in the Table \ref{NMT} at $\rho_0$ is in between 30.2 -- 33.7 MeV and 35.0 -- 70.0 MeV respectively. With the addition of DM, the values of $J$ and $L$ lie (except for NL3) in this region as given in the Table \ref{NMT}.\\
    
\begin{figure}
        \includegraphics[width=0.9\columnwidth]{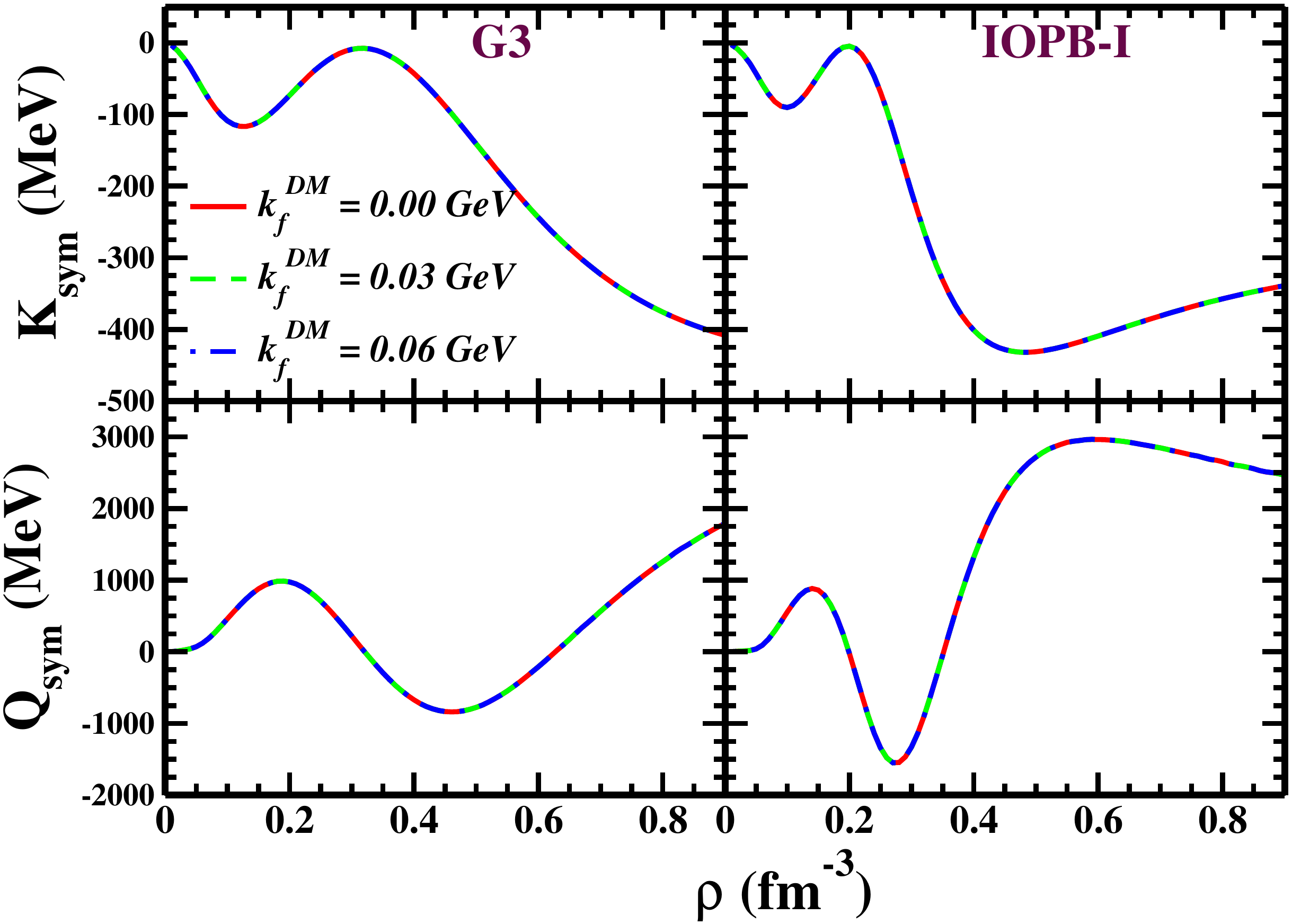}
        \caption{(colour online) The variation of $K_{sym}$ and $Q_{sym}$ with baryon density is plotted for G3 and IOPB-I at different  $k_f^{DM}$. The $K_{sym}$ and $Q_{sym}$ are opposite to each other both for G3 and IOPB-I.}
        \label{KQSYM}
    \end{figure}

    The other higher-order derivatives of symmetry energy like $K_{sym}$ and $Q_{sym}$ are also calculated in this section. The results are displayed in Fig. \ref{KQSYM} and their numerical values are tabulated in Table \ref{NMT}. The $K_{sym}$ is a parameter, which tells a lot not only about the surface properties of the astrophysical object (such as NS and white dwarf), but also the surface properties of finite nuclei. The whole density range of $K_{sym}$ and $Q_{sym}$ for G3 and IOPB-I sets are shown. The $K_{sym}$ affected marginally, but the parameter $Q_{sym}$ influenced significantly by DM (see Table \ref{NMT}) for these values at saturation. At low density, $K_{sym}$ initially decreases slightly, then it increases up to $\rho$ ($\sim$ 0.2 fm$^{-3}$) and after that decreases the value almost like an exponential function. Recently the value of $K_{sym}$ is calculated by Zimmerman et al.  \citep{Zimmerman_2020} with the help of NICER \citep{Miller_2019,Riley_2019,Bogdanov_2019,Bilous_2019,Raaijmakers_2019,Guillot_2019} and GW170817 data and its values lies in the range -174 to -31 MeV as given in Table \ref{NMT}. Our $K_{sym}$ values lie in this region at the saturation density.
    \subsection{Neutron Star Matter}\label{NS}
\begin{figure}
        \includegraphics[width=1.0\columnwidth]{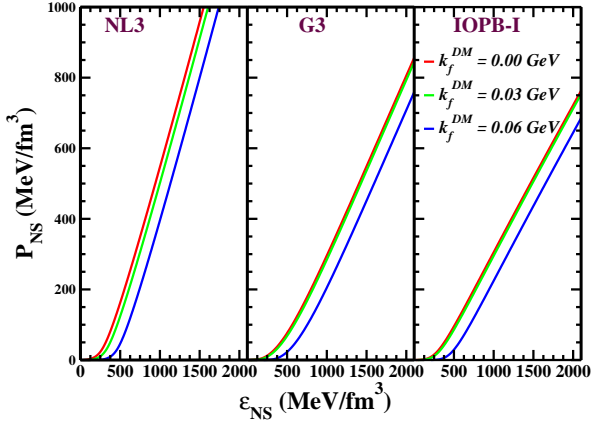}
        \caption{(colour online) The variation of energy density and pressure in Eq. (\ref{eq23}) with different $k_f^{DM}$. We find NL3 is the stiffest and G3 is the softest.}
        \label{EoSDM}
    \end{figure}

    As an application of our mixed EoS, i.e. DM and hadron matter, we used the EoS to calculate the M, R and I of RNS. For this, we constructed the star EoS maintaining the charge neutrality and $\beta$-equilibrium condition varying the DM momentum $k_f^{DM}$. The EoS for NL3, G3 and IOPB-I at $k_f^{DM}$= 0.0, 0.03 and 0.06 GeV are shown in Fig. \ref{EoSDM}. As we have mentioned earlier in the PNM and SNM, the EoS is very sensitive to $k_f^{DM}$, here also the energy density ${\cal{E}}$ is equally sensitive to the DM.  We find softer EoS  as in \citep{AngLi_2012,Panotopoulos_2017,Bhat_2019,Das_2019,Quddus_2019} with $k_f^{DM}$. Similar to the SNM and PNM, NL3 predicts the stiffest and G3 the softest EoS for different $k_f^{DM}$. The calculations of $M$, $R$ and $I$ of the SNS and RNS need the entire EoS of the NS. In crust region, which has very low density ( $< 5\times10^{14}$ g cm$^{-3}$ ), the inter nucleon distances is higher than the core region of the NS and the clusterisation of the nucleon happens. In the core region, due to the high density the NS system behave like a incompressible fluid of the nucleons. For the core part, we use the RMF equation of state. In crust part, we use BPS \citep{BPS_1971} EoS. Construction of the BPS EoS is based on the minimisation of the Gibbs function and effects of the Coulomb lattice, which gives a suitable combination of the A and Z. Once we know the entire EoS, we can calculate the properties of the SNS solving TOV equations \citep{TOV1, TOV2} and RNS using RNS code \citep{NikolaosStergioulas_1999}. The brief formalism for RNS can be found in Sub-Sec. \ref{RNS}. \\
    
\begin{figure}
        \centering
        \includegraphics[width=0.9\columnwidth]{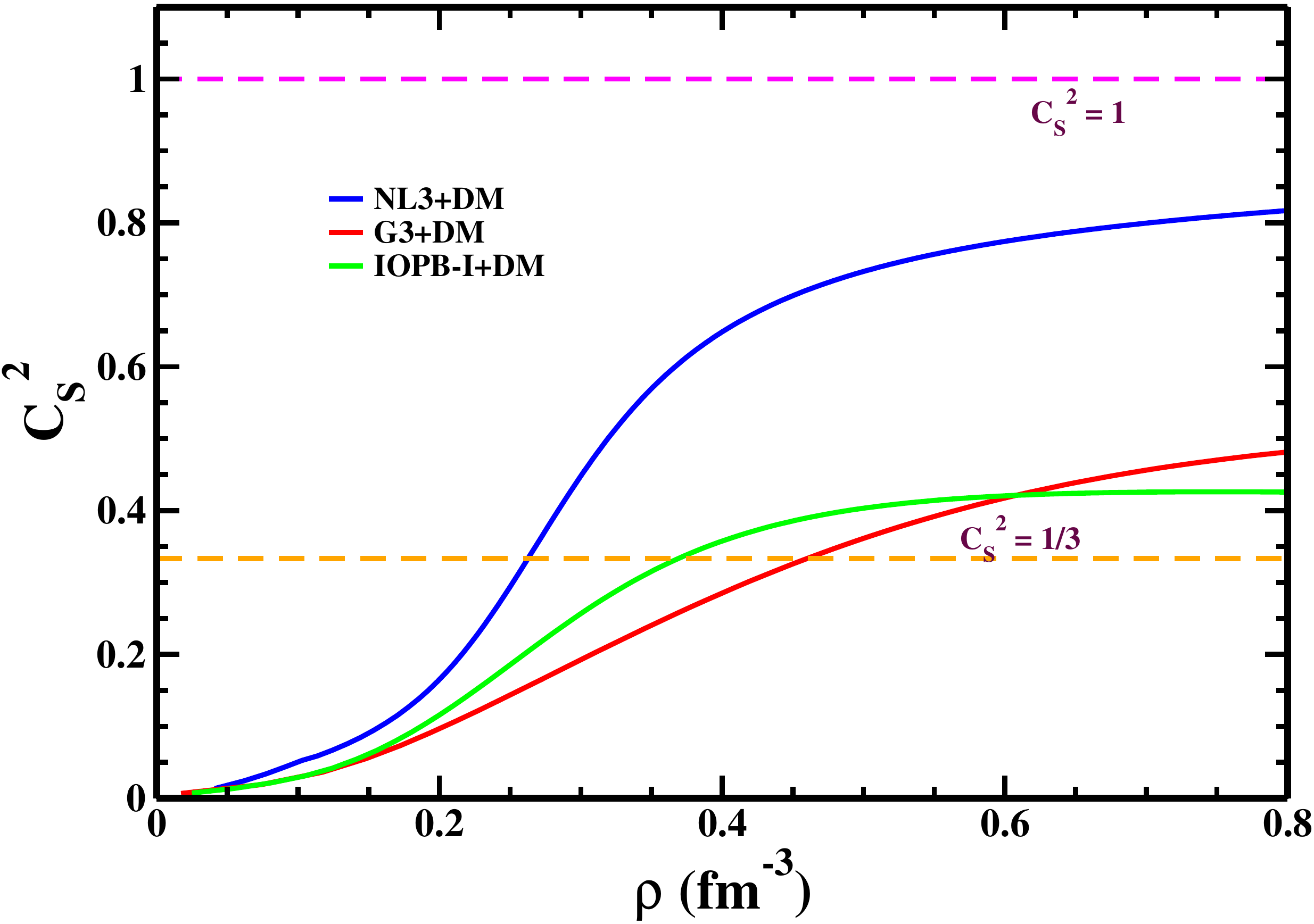}
        \caption{(colour online) The speed of the sound is plotted with the baryon density for 3-parameters sets with $k_f^{DM}$ = 0.06 GeV. The orange dashed line represents the conjecture (C$_S^2$ = 1/3) and magenta dashed line represents the C$_S^2$ is equal to c.}
        \label{VELDM}
    \end{figure}

    Before calculating the $M$ and $R$, we tested the causality condition \citep{Rhoades_1974,Bedaque_2015,Kojo_2015,Moustakidis_2017,McLerran_2019} in the NS medium with an admixture of DM. We have seen that both in NM and NS matter case, the causality is not violated throughout the region, as shown in Fig. \ref{VEL} and \ref{VELDM} respectively. The dashed horizontal line is the conjectured $C_s^2=1/3$ value given in Fig. \ref{VELDM}. The NL3 predicts the stiff rise in $C_s^2$ as compared to G3 and IOPB-I. But in all the cases $C_s^2$ is less than 1/3 for very low density region ($\rho <0.4$ $fm^{-3}$). As compared to the NM, the NS contains nucleons, electrons and muons, which is completely a different stable system survived by the balancing force due to the attractive gravitation and the repulsive degenerate neutrons also with short-range repulsive nuclear force. \\
    \begin{figure}
        \centering
        \includegraphics[width=0.9\columnwidth]{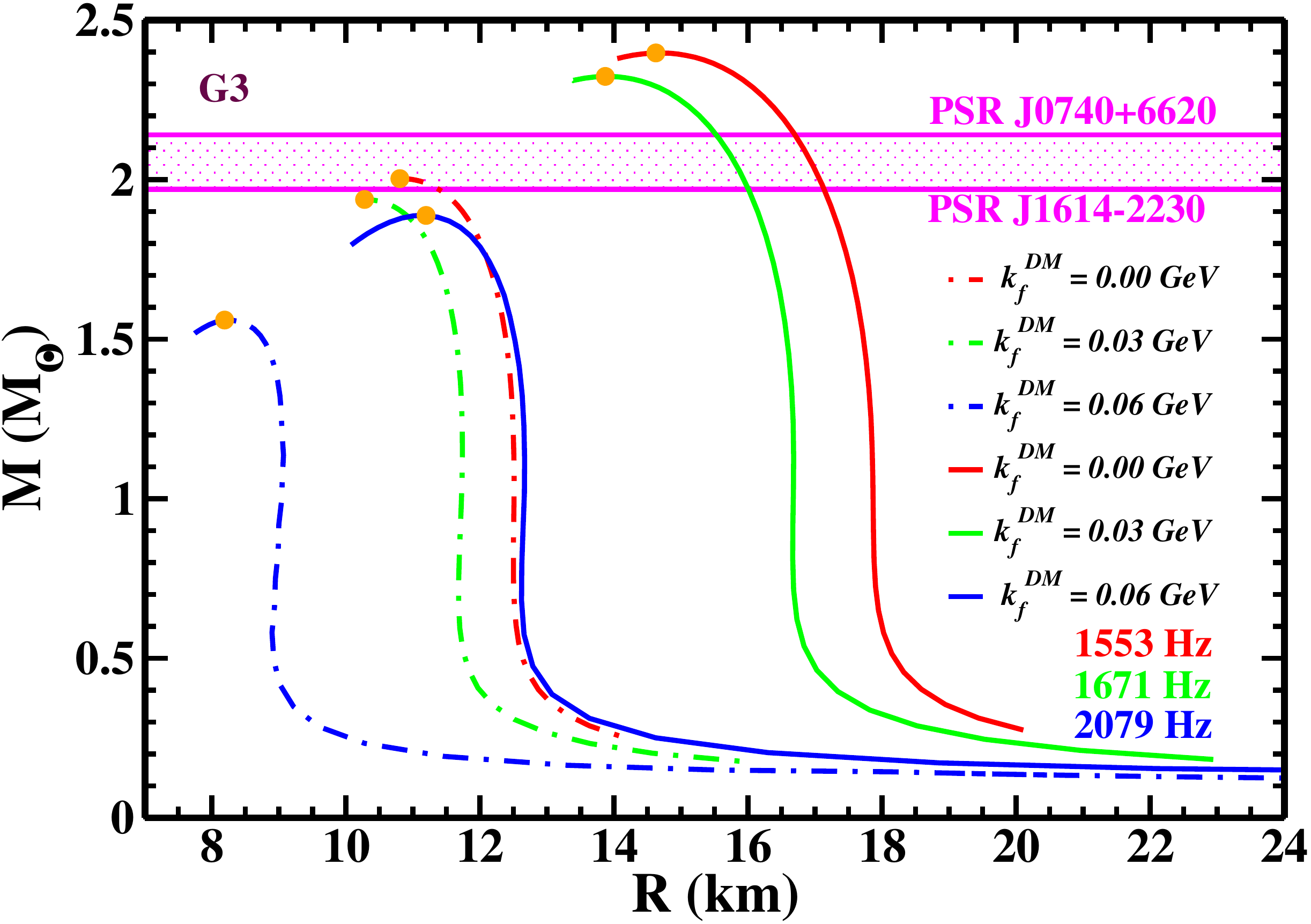}
        \caption{(colour online) The variation of mass with radius with different $k_f^{DM}$ is shown. The orange colour dot shows the maximum mass of the corresponding $k_f^{DM}$ for G3 parameter set. The dotted-dashed line represents for the SNS, and the bold line represents the RNS. The maximum rotational frequencies of NS also shown.  The recent observational constraints on NS masses \citep{Demorest_2010,Cromartie_2019} are also shown.}
        \label{MASS}
    \end{figure}

    Now, we calculate the M-R relations and $I$ for G3 as a representative case. The M-R relation for different k$_{f}^{DM}$ are shown in Fig. \ref{MASS}. The results from the precisely measured NSs masses, such as PSR J1614-2230 \citep{Demorest_2010} and PSR J0740+6620  \citep{Cromartie_2019} are shown in the horizontal lines in pink colours. These observations suggest that the maximum mass predicted by any theoretical model should reach the limit $\sim$ 2.0 $M_{\odot}$, and this condition is satisfied in all of the EoSs, which are taken into consideration. We noticed that the increase in $k_f^{DM}$, higher the energy density at a given $\rho$, which yield the lower the maximum mass and radius of the SNS and RNS. For RNS case the maximum mass is increased by $\sim$ 20\%, and radius increased by $\sim$ 26\% for the given EoS, which is approximately equal to the increase in mass due to the rapid rotation of the NS \citep{Stergioulas_2003, Worley_2008}.\\
    \begin{figure}
        \centering
        \includegraphics[width=0.9\columnwidth]{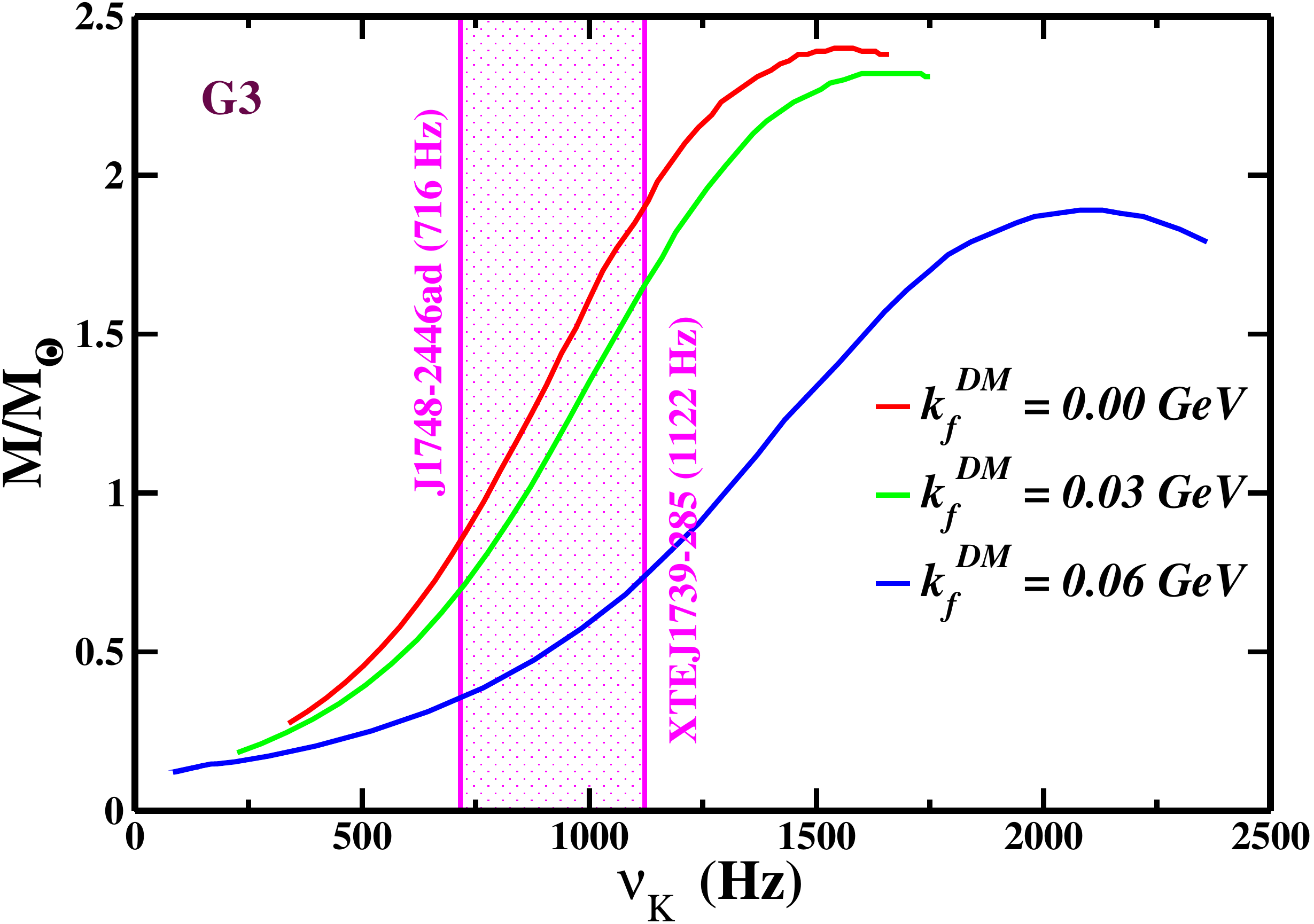}
        \caption{(colour online) The variation NS mass with Keplerian frequency $\nu_K$ is shown for G3 parameter set. The two vertical magenta line represents the frequencies of the fastest NSs J1748-2446ad \citep{Hessels_2006} and XTE J1739-285 \citep{Kaaret_2007}. }
        \label{FREQ}
    \end{figure}

    Here, we examine the effect of Kepler frequency ($\nu_K$) on the mass of the NS. In Fig. \ref{MASS}, the mass of the RNS is increased due to the rapid rotation of the NS due to its high $\nu_K$, i.e. the mass of NS directly depends on the $\nu_K$.  Theoretical calculations allows the value of  $\nu_K$ is more than 2000 $Hz$ \citep{Koliogiannis_2020}, but till now two fastest pulsar was detected having frequency 716 $Hz$ \citep{Hessels_2006} and 1122 $Hz$ \citep{Kaaret_2007}. From Fig. \ref{FREQ}, one can conclude that the NS mass approximately more than 1.7 $M_{\odot}$ is rotated more than the fastest pulsar as of today. In our case, we find the $\nu_K$s are 1553, 1671 and 2079 $Hz$ for the DM momentum 0, 0.03 and 0.06 GeV respectively at the maximum mass. The spherical NS is leading to deformed shape with the increasing of mass (or frequency)  \\
    
\begin{figure}
        \centering
        \includegraphics[width=0.9\columnwidth]{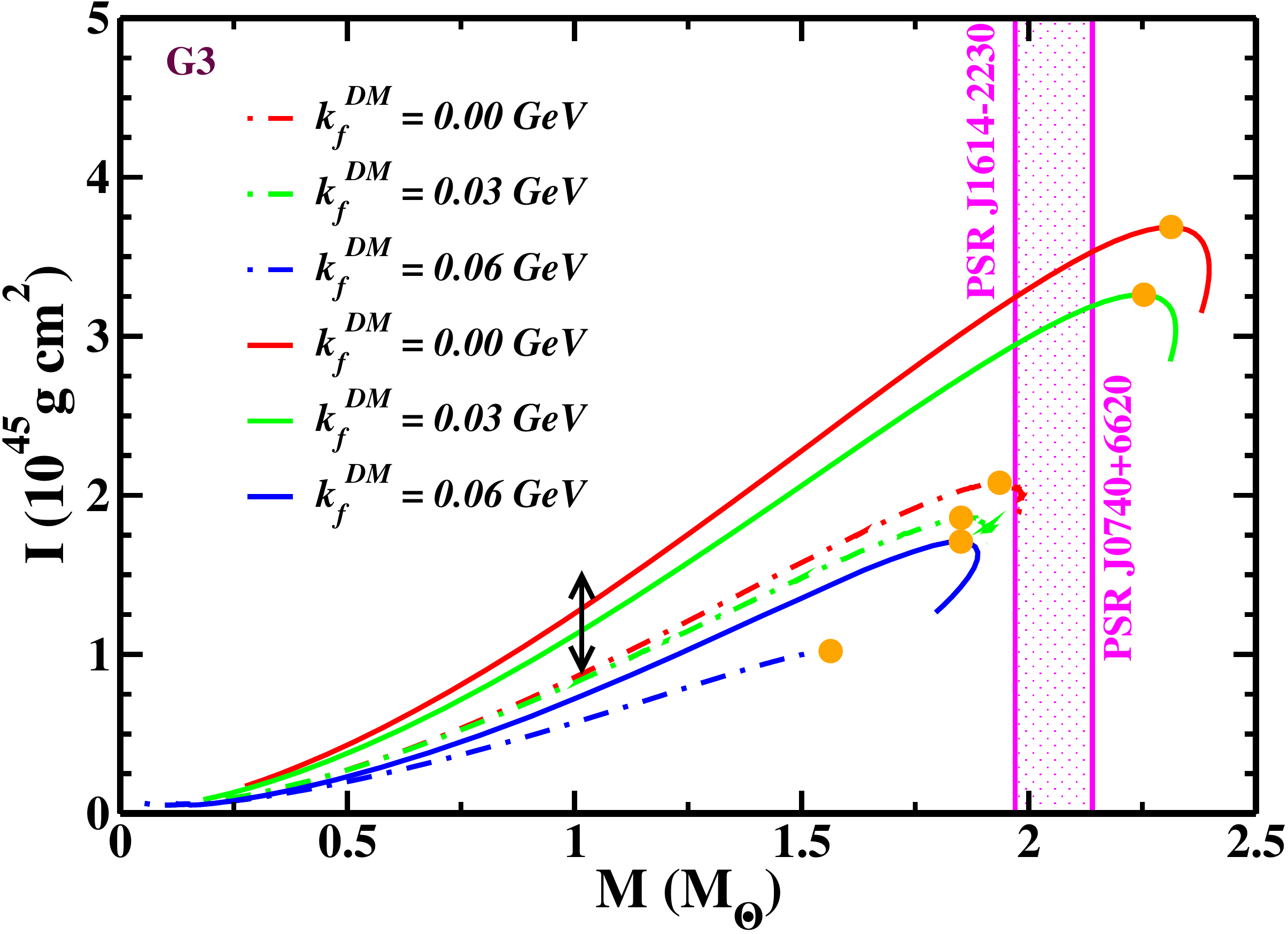}
        \caption{(colour online) The variation of $I$ with the mass of NS for different $k_f^{DM}$ using G3 parameter set. The orange colour dot shows the maximum mass of the corresponding $k_f^{DM}$. The dotted-dashed line represents for the SNS, and the bold line represents the RNS. The overlaid arrows represent the constraints on $I$, of PSR J0737-3039A set \citep{Landry_2018,Kumar_2019} from the analysis of GW170817 data \citep{Abbott_2017,Abbott_2018}}
        \label{MOM}
    \end{figure}

    A measurement of the $I$ of PSR J0737-3039A is expected via optical observation of the orbital precision in the double pulsar system in the near future \citep{Burgay_2003}. As the $I$ depends on the internal structure of the NS, it's measurement will constrain the unknown EoS of supra-nuclear matter, which is believed to be universal for all NS \citep{Landry_2018, Kumar_2019}. Here we show the variation of $I$  with $M (M_\odot$) in Fig. \ref{MOM}. The change of $I$ with $M$ is almost linear for different values of $k_f^{DM}$ up to the maximum mass of the star. Then there is a drop of $I$, as shown in Fig. \ref{MOM}.
    Since the increase of DM momentum ($k_f^{DM}$) leads to softer EoS, and hence the decrease of $I$. This is because $I$ ($\sim$ MR$^2$) is directly proportional to the mass and square of the radius of the rotating object. The moment of inertia is larger for the stiffer EoS as it predicts a larger radius and vice-versa.
    \section{Summary and Conclusions}\label{CONCLU}
    In summary, we studied the effects of DM on the NM parameters, such as nuclear incompressibility ($K$), symmetry energy ($S$) and its higher-order derivatives like $L$-slope parameter, $K_{sym}$-isovector incompressibility and $Q_{sym}$-skewness parameter for different asymmetric. These are the significant quantities responsible for the behaviour of nuclear EoS. The EoS becomes softer or stiffer depending on the values of these parameters. We calculated these quantities taking different admixture of DM fraction in the NM with varying neutron-proton asymmetry. We take Neutralino as a DM candidate which is trapped inside the NS and its interaction to nucleons through SM Higgs via Yukawa potential. The RMF Lagrangian with NL3, G3 and IOPB-I forces are used to get the hadron EoS, and the DM part is added on top of it.\\
    
    We find softer EoS with the increasing DM momentum, i.e. the energy density increases with $k_f^{DM}$ without adding much to the pressure. The influence of DM on effective mass, symmetry energy, $L$-coefficient and $K_{sym}$ are not much change with the variation of $k_f^{DM}$ due to the small contributions of the Higgs field. However, some other derivatives of $S$ (Q$_{sym}$) and $\cal{E}$ ($K$) affected significantly by DM. These effects contribute to the mass, radius and moment of inertia of stellar object like a NS. Also, the variation of ${\cal{E}}$, $K$ and $Q_{sym}$ due to DM not only affect the structure of NS but also significantly influence on the cooling effect of newly born NS after a supernova explosion. Thus, a detailed study of the temperature-dependent EoS is due, which will be published somewhere \citep{Kumar_2020}.\\
    
    To check the influence of DM on the accreting object, we constructed the NS EoS at various momenta of DM admixture. The mass, radius and the moment of inertia are evaluated for static as well as rotating cases using the TOV and RNS equations. The mass and radius are significantly reduced with the increase of DM momentum as we know that the stiffer EoS gives higher mass, radius and $I$ of a RNS. The mass of the NS is significantly changed due to the rapid rotation of the NS. Quantitatively, the mass of NS approximately more than 1.7 $M_{\odot}$ are rotating having the frequency more than the fastest pulsar ever detected. \\
    
    {\bf ACKNOWLEDGEMENTS}\\
    
    H. C. Das would like to thank T. K. Jha for the fruitful discussion about the RNS code and Abhishek Roy for the necessary discussion on the DM. This work is supported in part by JSPS KAKENHI Grant No.18H01209, and also  S. K. Biswal and Ang Li are supported by the National Natural Science Foundation of China Grant No. 11873040.
    
    \bibliographystyle{mnras}
    \bibliography{dm}
\end{document}